\documentclass[11pt,twoside]{article}
\usepackage{./asp2014}

\aspSuppressVolSlug
\resetcounters

\def\ab{Astrophys. Bull.}

\def\azh{Astron. Rep.}

\begin{document}

\title{Spectroscopic Study of the High-latitude  far Evolved Star V534\,Lyr}

\author{E.G.~Sendzikas,   
\affil{Special Astrophysical Observatory RAS, Nizhnij Arkhyz, Karachai-Cherkessia, Russia; email:{esendzikas@yandex.ru}}}
\author{E.L.~Chentsov,   
\affil{Special Astrophysical Observatory RAS, Nizhnij Arkhyz, Karachai-Cherkessia, Russia; email:{echen@sao.ru}}}

\paperauthor{Eugene  Sendzikas}{esendzikas@yandex.ru}{ORCID_Or_Blank}{Special Astrophysical Observatory RAS}{Astrospectrosopy laboratory}{Nizhnij Arkhyz}{Karachai-Cherkessia}{369167}{Russia}
\paperauthor{Eugene  Chentsov}{echen@sao.ru}{ORCID_Or_Blank}{Special Astrophysical Observatory RAS}{Astrospectrosopy laboratory}{Nizhnij Arkhyz}{Karachai-Cherkessia}{369167}{Russia}

\begin{abstract}
We study a pulsating variable post-AGB star  V534\,Lyr\,=\,HD172324 based on five high resolution spectra (R\,=\,60\,000) 
obtained with  the NES echelle spectrograph of  the 6--meter Russian telescope (BTA) in 2010 and 2013. 
Using the atmosphere modeling method and the Kurucz model set,  we obtained the effective temperature T$_{\rm eff}$\,=\,10500\,K, 
surface gravity  log\,g\,=\,2.5, and microturbulent velocity ${\rm \xi_t}$=4.0\,km/s. The underabundance of the 
iron group elements ${\rm [Met/H]_{\odot} = -0.50}$ was detected. This fact in combination with  
high spatial velocity indicates that V534\,Lyr does not belong to the disk population. The radial velocity gradient 
in the V534 Lyr atmosphere is  minimum: differential  shifts of lines are  close to measurement errors
The spectral  class A0\,Iab corresponds  to the distance to  V534 Lyr, d$\approx$6\,kpc.
\end{abstract}

\section{Introduction}\label{intro}

Over the two past decades AGB and post-AGB supergiants and several luminous stars with unclear evolutionary 
status have been spectroscopically monitored with the 6-m telescope of the Special Astrophysical Observatory 
(author of the program -- V.G.~Klochkova). Main results of this long-term study are presented by \citet{K2014, K2016}.

In the present paper we consider some details of the optical spectra of one of the objects of this program -- far 
evolved star  V534\,Lyr\,=\,HD\,172324.   V534\,Lyr  is a pulsating variable post-AGB star with the galactic coordinates: 
l\,=\,66.2$^{\rm o}$, b\,=\,18.6$^{\rm o}$.  Its spectral class A0\,Iab was determined by \citet{Bonsack}, physical 
properties and  chemical composition of the atmosphere are studied by \citet{Giridhar}.

This work is based on five high resolution spectra  (R\,=\,60\,000) obtained with the 
NES echelle spectrograph \citep{NES} of the 6-m Russian telescope (BTA) in 2010 and 2013. 
Extraction of one-dimensional spectra from two-dimensional echelle images was conducted with the modified 
variant by \citet{MIDAS} of the ECHELLE context of the MIDAS package. Wavelength calibration 
was carried out from the spectra of a Th--Ar hollow-cathode lamp.  One-dimensional spectra were reduced in the 
DECH20t code \citep{DECH}. 
Radial velocities were measured through fitting of the direct and mirror  images of line profiles. The fitting 
of spectra of the star and the lamp  was checked using the [OI], O$_2$, and H$_2$O telluric lines. The root-mean-square error 
of the radial velocity Vr for narrow absorptions is 1.0\,km/s at most.

\section{Main results}\label{results}

The table shows the data on the spectra and average heliocentric  velocities  Vr 
for several groups of lines. Columns 3--6 contain the averaged Vr for:
\begin{itemize}
\item FeII~6318, 6384, and 6385\,\AA{} emissions,  
\item HeI and S\,II absorption core,
\item components' core of the FeII absorptions of high  excitation (potentials of lower levels are equal to about 10 eV),
\item components' core of the FeII absorptions of low excitation (potentials of lower levels are equal to $\approx $1\,eV) 
respectively.
\end{itemize}

\begin{table}[!ht]
\smallskip
\begin{center}
\caption{The data on the spectra and average heliocentric  velocities  Vr in km/s for several groups of lines. } 
\medskip
\begin{tabular}{c c  c  c  c c }
\hline
Date     & $\Delta \lambda$  &Vr(emiss)&\multicolumn{3}{c}{Vr(absorptions)}\\
\cline{4-6}
\noalign{\smallskip} 
          &      nm         & FeII    &HeI, SII     & FeII high  & FeII low  \\         
\noalign{\smallskip}         
\hline 
      06.04.10 & 516-669    & $-131$  &  $-131$  &  $-132$  &$-103$   \\
               &            &         &          &          &$-152$   \\
\noalign{\smallskip}                                                                             
\hline
      01.06.10 & 522-669    &  $-123$ & $-120$   & $-120$   & $-107$  \\
               &            &         &          &          & $-156$  \\
\noalign{\smallskip}
\hline               
30.07.10       & 443-593    &         &  $-124$  & $-125$   & $-125$  \\
\noalign{\smallskip}
\hline
24.09.10       & 522-669    &  $-127$ &  $-130$  &  $-130$  & $-128$  \\
\noalign{\smallskip}
\hline
12.10.13       & 392-698    &  $-114$ & $-120$   &  $-108$  & $-109$  \\
               &            &         &          &          & $-146$  \\
\noalign{\smallskip}                                                                           
\hline 
\end{tabular}  
\label{Spectra}  
\end{center}  
\end{table}

The figure presents the dependences between the radial velocity Vr and the central 
residual  intensity (r). Each line corresponds to one or two signs (in cases of  bifuricated absorptions). 
As distinct from  other super and hypergiants, for this star they are not stationary and show velocities 
close to those for unsplit absorptions. 

The table and figure  obviously show the position variations of all 
the lines and shapes  of profiles of some of them with time. This primarily  refers to the splitting of FeII 
absorptions of low excitation forming in the upper atmospheric layers; the absorptions of high excitation forming 
deeper stay single. 

\begin{figure}[!ht]
\includegraphics[angle=90,width=6cm,height=21cm,bb=8 20 750 215,clip]{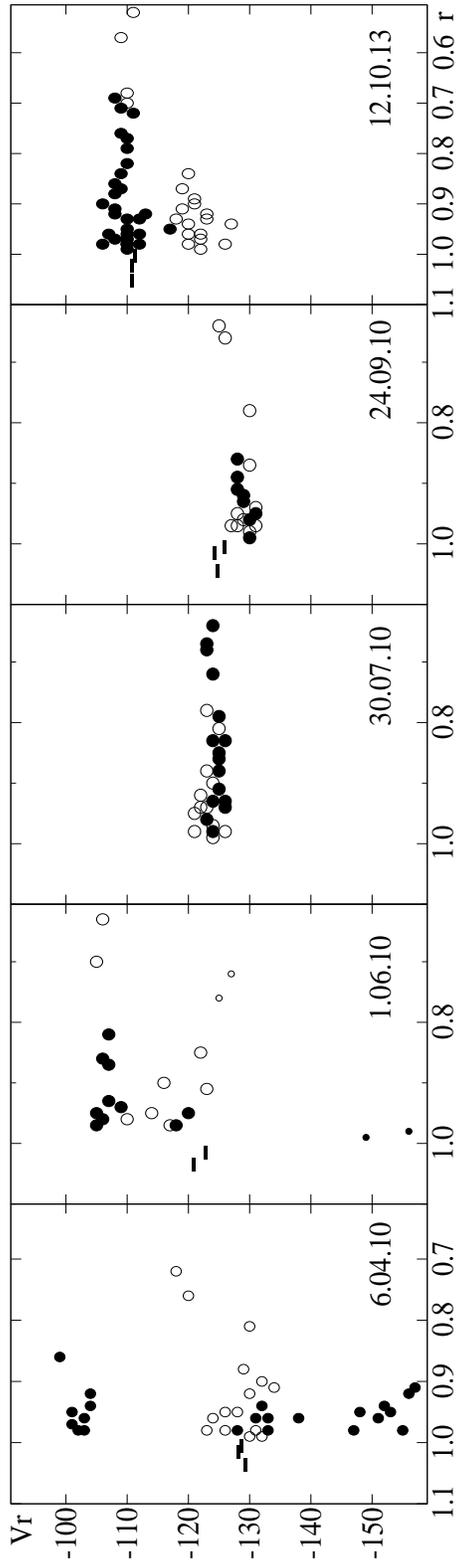}
\caption{Relation between the radial velocity Vr and the central residual  intensity (r).  
         The  filled circles refer to FeII, the open ones  -- to HeI, SII, and SiII~(2), horizontal little lines -- 
         to emissions. }
\label{Vr-pattern}
\end{figure}

The radial velocity gradient in the V534\,Lyr atmosphere is  minimum, 30.07.10 and 24.09.10: differential 
shifts of lines are  close to measurement errors. On that basis, one can suppose that 
center-of-mass velocity of the star, Vsys,  is close to $-125$\,km/s (V$_{\rm lsr}\approx -105$\,km/s). 

The spectral class A0\,Iab corresponds to the distance to  V534\,Lyr, d$\approx6$\,kpc. 
Interstellar line profiles of NaI~(1) and CaII~(1)  are also indicative of great distance: 
the presence of the components with Vr\,=$-46$\,km/s in them indicates  d$>7$\,kpc as in 
\citep{Brand}. Notable proper motion (3.6$\pm$0.8\,mas) at d$\approx6$\,kpc  is corresponding 
to 103\,km/s; the star is near to the Galactic plane. In combination with V$_{\rm lsr}\approx -105$\,km/s, 
it yields a spatial velocity of  about 140\,km/s. 

Using the model atmospheres method and the models set by \citet{models}, we obtained the effective 
temperature T$_{\rm eff}$\,=\,10500\,K, surface gravity  log\,g\,=\,2.5, and microturbulent velocity 
${\rm \xi_t}$\,=\,4.0\,km/s. 
The underabundance of the  iron group elements   ${\rm [Met/H]_{\odot} = -0.50}$
was also calculated with this atmospheric model. This fact as well as  high spatial velocity indicates that 
V534\,Lyr does not belong to the disk population. 

Among stars  we studied earlier, the anomalous supergiant UU\,Her, the prototype of the class of variable 
supergiants located at high galactic latitudes  may be considered  as a nearest analogy of  V534\,Lyr.  
Based  on  several high-resolution  spectroscopy of the star over 5 years with the 6--m telescope, 
\cite{UUHer} found a low metallcity ${\rm [Fe/H]_{\odot} = -1.32}$  and the large radial velocity  
${\rm V_r \approx 130}$\,km/s.  In whole  obtained results led these authors to conclude that UU\,Her 
is a low-mass  star beyond Galactic disk.

\acknowledgements 
This study was accomplished with a financial support of the Russian Foundation for Basic Research  
in the framework of the project  No.\,14--02--00291\,a.

\end{document}